# Tuning Electrode Wettability to Optimize Nanobubble Nucleation and Reaction Rates in Electrochemical Gas-Evolving Reactions

Zhenlei Wang[a,b,c], Yaxi Yu[a,b,c], Mengkai Qin[a,b,c], Hao Jiang[a,b,c], Zhenjiang Guo[b], Lu Bai[a], Limin Wang[b,c], Xiaochun Zhang[a,b] *, Xiangping Zhang[a,b] *, Yawei Liu[a,b,c,d] *

[a]*Beijing Key Laboratory of Solid State Battery and Energy Storage Process, CAS Key Laboratory of Green Process and Engineering, Institute of Process Engineering, Chinese Academy of Sciences, Beijing 100190, P. R. China*
[b]*State Key Laboratory of Mesoscience and Engineering, Institute of Process Engineering, Chinese Academy of Sciences, Beijing 100190, P. R. China*
[c]*University of Chinese Academy of Sciences, Beijing 100049, P. R. China*
[d]*Longzihu New Energy Laboratory, Zhengzhou Institute of Emerging Industrial Technology, Henan University, Zhengzhou 450000, P. R. China*
*E-mail: xchzhang@ipe.ac.cn, xpzhang@ipe.ac.cn, ywliu@ipe.ac.cn

**TOC**

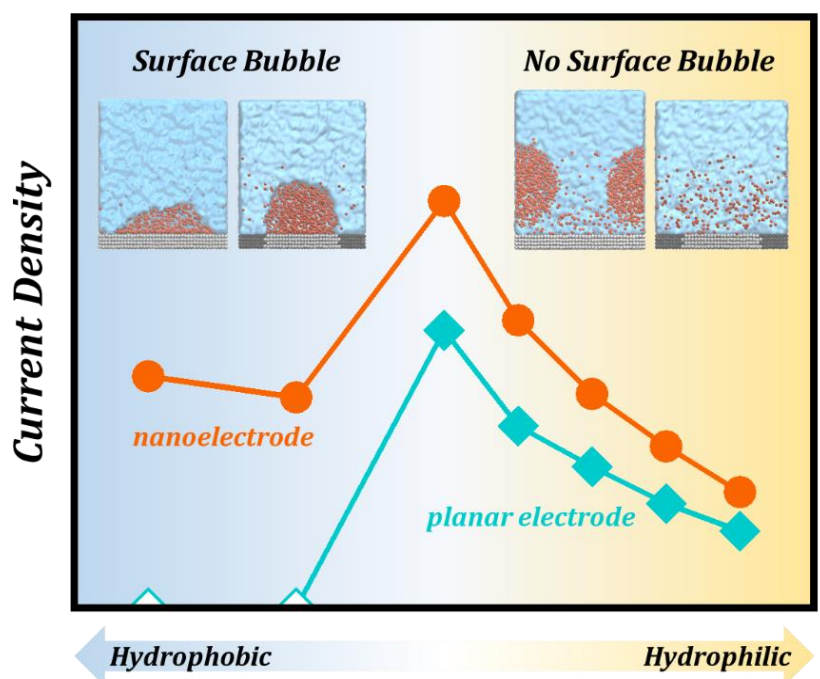

**Abstract**

Bubble formation in electrochemical system often hinders reaction efficiency by reducing active surface area and obstructing mass transfer, yet the mechanisms governing their nanoscale nucleation dynamics and impact remains unclear. In this study, we used molecular dynamics simulations to explore nanobubble nucleation and reaction rates during water electrolysis on planar- and nano-electrodes, with systematically tuning electrode wettability through water-electrode and gas-electrode interactions. We identified distinct nucleation regimes: gas layers, surface nanobubbles, bulk nanobubbles, and no nanobubbles, and revealed a volcano-shaped relationship between wettability and reaction rate, where optimal wettability strikes a balance between suppressing bubbles and ensuring sufficient reactant availability to maximize performance. Nanoelectrodes consistently exhibit higher current densities compared to planar electrodes with the same wettability, due to pronounced edge effects. Furthermore, moderate driving forces enhance reaction rates without triggering surface bubble formation, while excessive driving forces induce surface nanobubble nucleation, leading to suppressed reaction rates and complex dynamics driven by bubble growth and detachment. These findings highlight the importance of fine-tuning wettability and reaction driving forces to optimize gas-evolving electrochemical systems at the nanoscale and underscore the need for multiscale simulation frameworks integrating atomic-scale reaction kinetics, nanoscale bubble nucleation, and microscale bubble dynamics to fully understand bubble behavior and its impact on performance.

1. Introduction

Bubble formation is a ubiquitous phenomenon in electrochemical reactions, spanning large-scale industrial electrochemical processes such as the chlor-alkali process, sodium chlorate production, and aluminum production[1-2], as well as emerging technologies like water electrolysis[3-8], $CO_2$ electroreduction[9-12], and organic electrosynthesis[13-14]. Once bubbles form on the electrode surface, they reduce the active area available for reactions, thereby hindering electrochemical performance[15-16]. Additionally, bubbles near the electrode obstructs ion diffusion pathways, which decreases charge transfer efficiency[17-18]. Addressing the negative impact of bubbles is essential for improving reaction efficiency and can be approached through strategies such as suppressing bubble formation, encouraging bulk nucleation (i.e., bubble formation in the bulk solution rather than on the electrode surface), or facilitating bubble growth and detachment.

Recent experiments have demonstrated that carefully designing and optimizing electrode wettability can effectively promote bubble detachment, enhance mass transfer, and improve electrocatalytic performance[19-22]. However, most of these studies have focused on the macroscopic behavior of bubbles (typically >10 μm), leaving a significant knowledge gap regarding the formation and dynamics of nanobubbles at the electrode interface in these experiments. This is particularly critical because nanobubbles exhibit properties and behaviors that are markedly different from those of their macroscopic counterparts[23-29], potentially exerting a more fundamental influence on electrochemical reaction mechanisms at the nanoscale.

Advances in nanoscale characterization techniques have provided valuable insights into nanobubble evolution and their impact on electrochemical processes. For example, Zhang et al.[30-31] utilized cyclic voltammetry to detect the generation of a single CO nanobubble on a gold nanoelectrode during $CO_2$ electroreduction, yielding detailed information on nucleation parameters such as critical concentrations, nucleation rates, energy barriers, and bubble morphologies. Similarly, Chen et al.[32-33] employed scanning electrochemical cell microscopy combined with single-bubble nucleation voltammetry to explore nanobubble nucleation on silicon nanoparticles, highlighting the role of surface topology in heterogeneous nucleation. Despite these advances, real-time, *in situ* observations of individual nanobubble nucleation events and their associated electrochemical responses remain challenging due to limitations in spatial and temporal resolution as well as difficulties in controlling bubbles[34-37]. Moreover, experimental efforts to tune electrode wettability often introduce complex coupled effects, simultaneously altering intrinsic reaction kinetics and nanobubble behaviors, thereby complicating the identification of underlying mechanisms and causality.

Molecular simulation techniques offer a powerful complementary approach for overcoming these limitations. By enabling precise and independent control over system parameters, simulations provide real-time insights into molecular dynamics and nanobubble behavior in electrochemical systems. Recent simulation studies have been instrumental in advancing the understanding of nanobubble dynamics. For example, Sirkin et al.[38] investigated nanobubble nucleation and steady-state mechanisms on nanoelectrodes, achieving strong agreement with experimental observations of critical cluster formation. Zhang et al.[39] demonstrated the influence of gas solubility and reactant concentration on nanobubble stability and dynamics, while Lohse et al.[40] systematically examined the effects of current density on contact angle and bubble stability, further refining theoretical models of surface nanobubbles.

Despite these achievements, a fundamental question persists: how can electrode wettability be systematically optimized to regulate nanobubble nucleation and enhance reaction rates at the nanoscale? Resolving this question is crucial for advancing gas-evolving electrochemical systems and achieving precise control over bubble dynamics at the microscopic level.

In this work, we employed molecular dynamics simulations to investigate nanobubble nucleation during water electrolysis on both planar- and nano-electrodes (Figure 1 and Sec. Methods). By systematically tuning electrode wettability through varying water-electrode (W-E) and gas-electrode (G-E) interactions ($\varepsilon_{WE}$ and $\varepsilon_{GE}$), we identified distinct nanobubble nucleation regimes and constructed comprehensive phase diagrams to demonstrate how tailored wettability can regulate bubble nucleation behaviors. Our further analysis revealed a volcano-shaped relationship between electrode wettability and actual reaction rate, emphasizing the need for precise wettability tuning to suppress surface bubble formation and optimize electrochemical performance. We found that nanoelectrodes consistently achieve higher current densities than planar electrodes with equivalent wettability, primarily due to the pronounced influence of edge effects. Additionally, we investigated the role of the reaction driving force—quantified by the frequency $F$ of water-to-gas conversion attempts near reaction sites—and found that achieving an optimal balance of driving force is also critical for maximizing reaction efficiency at the nanoscale on electrodes with finely tuned wettability.

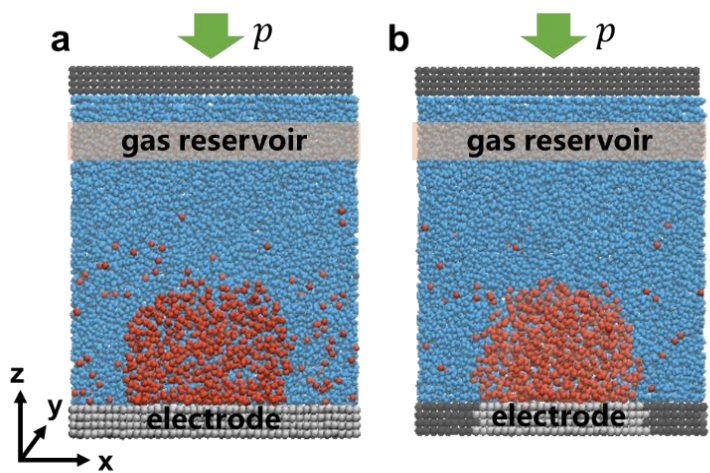

**Figure 1.** Typical simulation setups with (a) a planar electrode and (b) a nanoelectrode. Active Pt-like atoms are gray, inert Si-like atoms are black, water/reactant molecules are blue, and gas/product molecules are red (Sec. Methods).

## 2. Results and Discussion

### 2.1 Nanobubble Nucleation on Electrodes with Different Wettabilites

We first studied nanobubble nucleation on planar electrodes (Figure 1 a) with different wettabilities characterized by contact angles (CAs, Figure S1) at a moderate reaction driving force ($F = 0.1 \text{ ps}^{-1}$). As shown in Figure 2 a-d, with the electrode becomes more hydrophilic (i.e., CA decreases), four distinct regimes were observed sequentially: gas layers, surface nanobubbles, bulk nanobubbles, and no nanobubbles. Specifically, for highly hydrophobic electrodes (e.g., CA ~180° in Figure 2 a), the produced gas molecules are adsorbed onto the electrode surface, forming a dense gas layer that covers the entire electrode due to the relatively strong G-E attraction. This layer eventually reaches a steady state, where the rate of gas production balances the rate of gas diffusion away from the electrode (Movie S1). When the electrode becomes moderately hydrophobic (e.g., CA ~122° in Figure 2 b), gas molecules accumulate near the surface until nucleate into a surface nanobubble. Further increasing the electrode hydrophilicity (e.g., CA ~44° in Figure 2 c) results in bubble nucleation occurring just above the electrode, suggesting that the electrode is unable to promote heterogeneous nucleation due to the relatively weak G-E attraction compared to the W-E attraction. Notably, both surface and bulk nanobubbles on planar electrodes remain unstable, continuously growing and causing rapid expansion of the simulation domain

(Movie S1), implying that in experimental systems, such nanobubbles would continue to grow or merge with others to form micrometer-sized bubbles, which may eventually detach from the electrode surface due to buoyancy. Finally, for a highly hydrophilic electrode (e.g., CA ~ 0° in Figure 2 d), no bubble nucleation occurs throughout the simulation (Movie S1), indicating that the local gas concentration near the electrode never reaches the threshold for either heterogeneous or bulk nucleation.

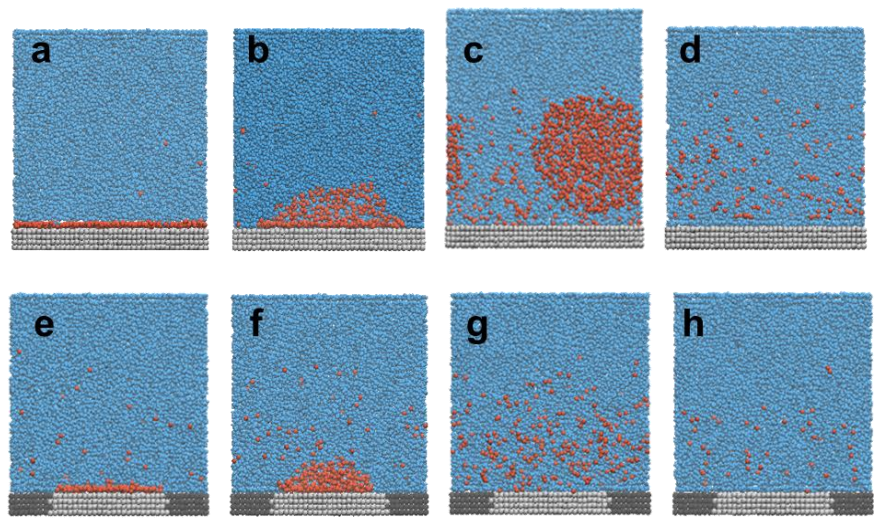

**Figure 2.** Nanobubble nucleation on (a-d) planar electrodes and (e-h) nanoelectrodes at varying wettabilities with (a, e) CA~180° ($\varepsilon_{WE}, \varepsilon_{GE} = 0.6, 1.0$ kcal/mol), (b, f) CA~122° ($\varepsilon_{WE}, \varepsilon_{GE} = 0.6, 0.6$ kcal/mol), (c, g) CA~40° ($\varepsilon_{WE}, \varepsilon_{GE} = 0.6, 0.2$ kcal/mol), and (d, h) CA~0° ($\varepsilon_{WE}, \varepsilon_{GE} = 1.0, 0.2$ kcal/mol).

Interestingly, the results in Figure 2 c-d suggest that the actual gas production rate, and consequently the gas supersaturation near the electrode, is influenced by the electrode wettability even at the same reaction driving force (i.e., the same $F$). Intuitively, one might expect that a more hydrophilic electrode, with higher water content at the surface, would facilitate a higher reaction rate. However, a closer examination revealed that while hydrophilic electrodes indeed attract more water molecules overall, the strong short-range repulsion introduced by large $\varepsilon_{WE}$ significantly reduces the density of water molecules in the immediate vicinity of the electrode—precisely where the reaction occurs (Figure S2). As a result, electrodes with a moderate hydrophilicity (i.e., a weak W-E interactions) exhibit a higher reaction probability (see Sec. 2.3 for more details), enabling more rapid gas accumulation near the electrode and thereby promoting nanobubble nucleation (e.g., Figure 2 c).

On 6-nm-diameter nanoelectrodes (Figure 1 b), similar trends to those observed on planar electrodes at extreme wettablities were observed: gas layers form on highly hydrophobic electrodes, while no nanobubbles form on highly hydrophilic electrodes (Figure 2 e and h). On moderately hydrophobic nanoelectrode (e.g., CA ~ 122° in Figure 2 f), surface nanobubbles are generated on the nanoelectrode. However, due to the contact line pinning effect, these nanobubbles are eventually stabilized in the gas supersaturated environment near the electrodes[26-27, 29, 41] (Movie S1). Under moderately hydrophilic conditions (e.g., CA ~ 40° in Figure 2 g), unlike on the planar electrode, no nanobubbles form on the nanoelectrode. This absence of bubble nucleation is attributed to the smaller effective reactive area of nanoelectrodes compared to planar electrodes, resulting in a lower gas supersaturation near the electrode, which is insufficient to drive bubble nucleation.

## 2.2 Phase Diagrams of Nanobubble Nucleation

To gain a comprehensive understanding of how nanobubble nucleation can be regulated through wettability modulated by W-E and G-E interactions, we systematically investigated nanobubble formation across various combinations of $\varepsilon_{WE}$ and $\varepsilon_{GE}$, and constructed the phase diagrams shown in Figure 3. For planar electrodes, as illustrated in Figure 2 a-d, increasing the electrode hydrophilicity (approximately from the top left to the bottom right in Figure 3 a) sequentially results in gas layers, *unstable* surface nanobubbles, *unstable* bulk nanobubbles, and no nanobubbles. Notably, bulk nanobubbles only emerge at weak G-E interactions (e.g. $\varepsilon_{GE} < 0.45$ in Figure 3), suggesting that the bulk nanobubble nucleation only occur when the adsorption capacity of electrode for gas molecules is insufficient to promote heterogeneous nucleation. At stronger G-E interactions, the electrode surface is more likely to facilitate heterogeneous nucleation, resulting in the formation of surface nanobubbles instead.

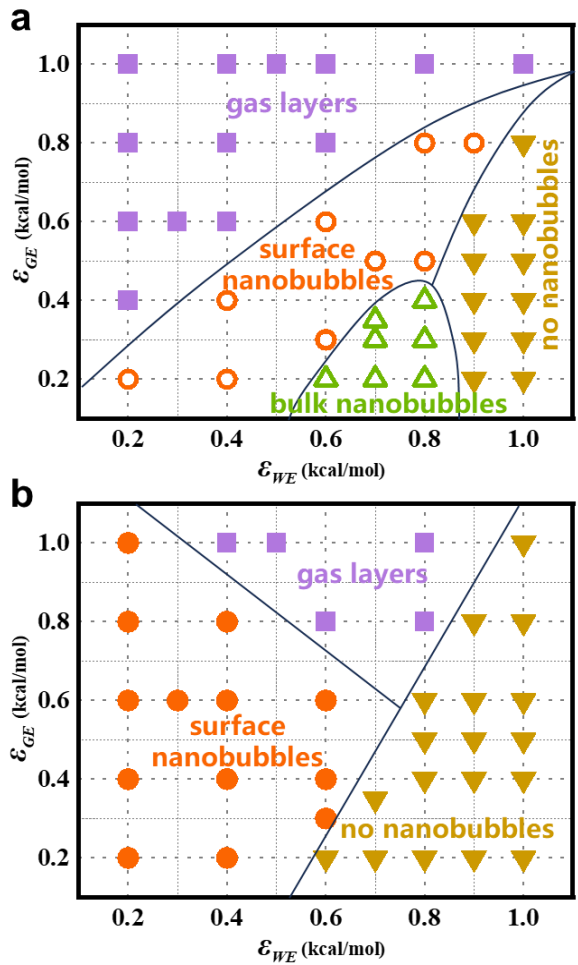

**Figure 3.** Phase diagrams of nanobubble nucleation on (a) planar electrodes and (b) nanoelectrodes at various water-electrode and gas-electrode interactions ($\varepsilon_{WE}$ and $\varepsilon_{GE}$). Solid symbols correspond to steady states, while hollow symbols denote unsteady states. The curves approximately indicate the boundaries between different regions.

In contrast, the nucleation phase diagram for nanoelectrodes shows significant differences compared to that of planar electrodes. The no-nanobubble region expanded considerably, which can be attributed to the reduced reactive area on nanoelectrodes, resulting in slower gas accumulation and lower gas supersaturation near the electrode. More importantly, we observe a substantial increase in the surface-nanobubble region on nanoelectrodes. Regions that manifest as gas layers on planar electrodes transition into stable nanobubbles on nanoelectrodes. To elucidate the underlying mechanisms, we analyzed the spatial distribution of reaction events under $\varepsilon_{WE} = 0.2$ and $\varepsilon_{GE} = 1.0$ at where the gas layers formed on the planar electrode while stable surface nanobubbles formed on the nanoelectrode. Our analysis revealed that gas molecules formed a dense adsorption layer on the electrode surface, blocking most reactions. However, thermodynamic fluctuations at the water-electrode-gas contact line intermittently exposed the electrode boundary, maintaining a relatively high reaction rate at the edges (Movie S2). This edge effect contributes to a faster gas production rate on gas-layer-covered nanoelectrodes compared to planar electrodes, facilitating the formation of nanobubbles (Figure S3 a-b). Interestingly, even in the absence of gas layers and surface nanobubbles, nanoelectrodes exhibit a pronounced edge effect, with a higher reaction probability occurring near the edges (Figure S3 c-d). We attribute this to the highly hydrophilic inert Si-like atoms near the edges, which leads to increased water density and reduced gas adsorption near the boundary, thereby enhancing the reaction likelihood in the narrow edges.

### 2.3 Influence of Wettability on Actual Reaction Rates

We next analyzed the actual reaction rates across different electrode wettabilities during simulations. Our simulations revealed that, except for cases involving surface nanobubbles on planar electrodes—where the system remains in an unsteady state until rapid expansion occurs—all other systems achieve steady states or at least experience a transient steady state regarding to the gas production rate (e.g., bulk nanobubbles on planar electrodes). To quantify the actual reaction rates in these steady states (Figure S4), the electric current was calculated based on the moles of gas produced per unit time, multiplied by the Faraday constant, assuming that two electrons are transferred per gas molecule. The current density is equated to the actual reaction rate. For a direct comparison between planar- and nano-electrodes, the current was normalized by the actual reactive area (which is 100 nm$^2$ for planar electrodes and 27 nm$^2$ for nanoelectrodes), yielding the *effective current density* ($J$).

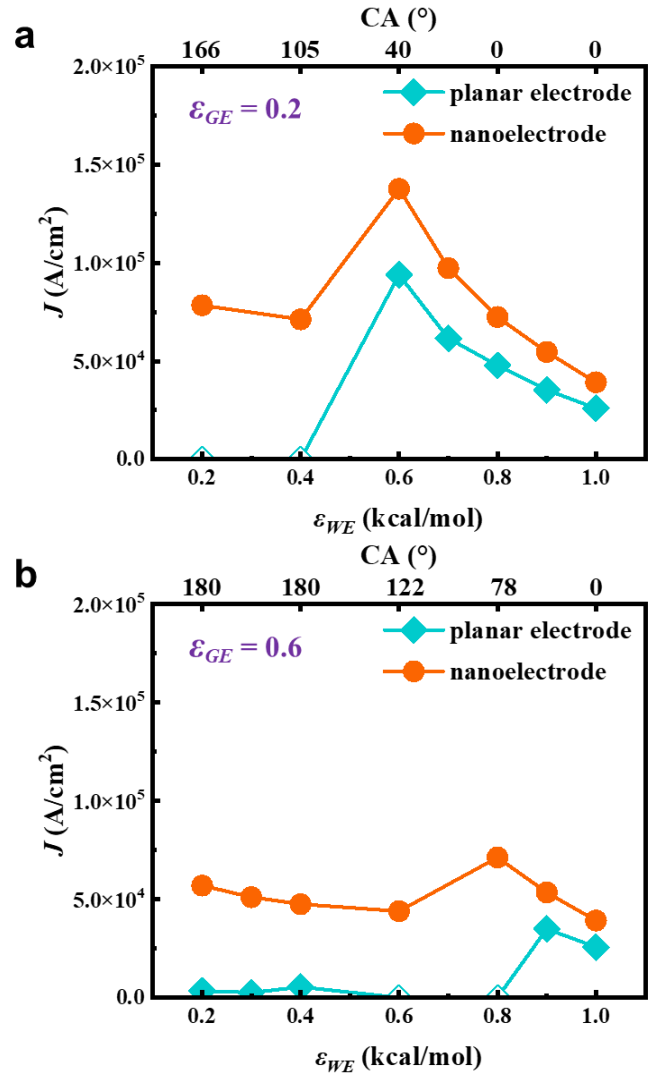

**Figure 4.** Effective current density $J$ on planar electrodes and nanoelectrodes at different water-electrode and gas-electrode interactions with (a) $\varepsilon_{GE} = 0.2$ kcal/mol; (b) $\varepsilon_{GE} = 0.6$ kcal/mol.

Figure 4 a illustrates the variation of $J$ with the wettability (i.e., CA) modulated by the W-E interaction (i.e., $\varepsilon_{WE}$) at a weak G-E interaction (i.e., $\varepsilon_{GE} = 0.2$ kcal/mol). In cases with unstable surface nanobubbles on planar electrodes, $J$ is set to zero to represent the scenario where the electrode surface remains fully obstructed for a significant duration, dictated by the growth dynamics and detachment intervals of the bubbles, resulting in an extremely low actual reaction rate. Interestingly, the relationship between $J$ and $\varepsilon_{WE}$ (as well as CA) follows a volcano-shaped trend, with the current density peaking at an optimal value of $\varepsilon_{WE}$. Specifically, when the W-E interaction is weak (e.g., $\varepsilon_{WE} < 0.6$), unstable surface nanobubbles form on planar electrodes, which is expected to result in a very low current density (e.g., $J\sim0$ in Figure 4 a). As the W-E interaction strengthens, the electrode's capacity to adsorb gas molecules diminishes, transitioning the nucleation regime from heterogeneous to bulk nanobubble nucleation or even no-nanobubble state. In the absence of bubbles covering the electrode, the actual reaction rate increases significantly and becomes primarily influenced by how the W-E interaction regulates the reaction kinetics in the reaction zone. In our model, larger $\varepsilon_{WE}$ correlates with lower reactant concentrations in the reaction region, reducing reaction probability and yielding lower effective current densities. Consequently, our simulations suggest that the optimal scenario for maximizing reaction efficiency involves precisely tuning the electrode wettability to prevent surface nanobubble formation while maintaining favorable reactant availability in the reaction zone.

A similar volcano-shaped relationship between $J$ and $\varepsilon_{WE}$ is also observed on nanoelectrodes, with the current density peaking at an optimal wettability (Figure 4 a). Furthermore, our nanoelectrode model exhibits enhanced reaction rates at the edges, resulting in an effective current density that surpasses that of planar electrodes with the same wettability. This finding suggests that nanostructuring electrodes can effectively improve the utilization efficiency of active materials, for example, reducing the required amount of catalyst to achieve the same current density. Such an approach is particularly meaningful for electrochemical reactions utilizing expensive noble metals as catalysts, such as hydrogen evolution reactions with Pt and oxygen evolution reactions with Ir[42-45].

The volcano-shaped relationship is also observed at stronger G-E interactions (e.g., $\varepsilon_{GE} = 0.6$ in Figure 4 b). However, increasing $\varepsilon_{GE}$ enhances the electrode's ability to adsorb gas molecules, requiring a higher $\varepsilon_{WE}$ to render the electrode sufficiently hydrophilic to prevent the formation of gas layers and surface nanobubbles. Moreover, a larger $\varepsilon_{GE}$ leads to a higher concentration of gas molecules near the electrode, reducing the overall reaction probability. As a result, the effective current density is lower compared to cases with lower $\varepsilon_{GE}$. Therefore, minimizing $\varepsilon_{GE}$ is generally beneficial for enhancing reaction efficiency.

**2.4 Influences of Reaction Driving Force**

The results above were obtained under a moderate reaction driving force. As a key parameter, the reaction driving force—related to the applied electrode voltage—governs the intrinsic reaction rate and the potential gas accumulation rate near the electrode surface, directly influencing bubble nucleation. In practical applications, even with finely tuned wettability to reduce surface bubble formation, bubbles may still arise due to factors such as operational fluctuations or additional nucleation sites caused by surface or bulk heterogeneity[46-48]. These bubbles, if adhered to the electrode surface, can significantly impede electrochemical reactions, particularly on nanoelectrodes where bubbles may stabilize[49-50], rendering certain reactive regions inactive for extended periods. In such cases, increasing the applied voltage to destabilize bubbles, promote their growth, and facilitate their detachment emerges as a practical solution to enhance gas production rates. To further investigate the influence of the reaction driving force, we then investigated bubble nucleation behavior and the corresponding changes in current density under two additional reaction driving forces ($F = 0.05$ and $1$ ps$^{-1}$).

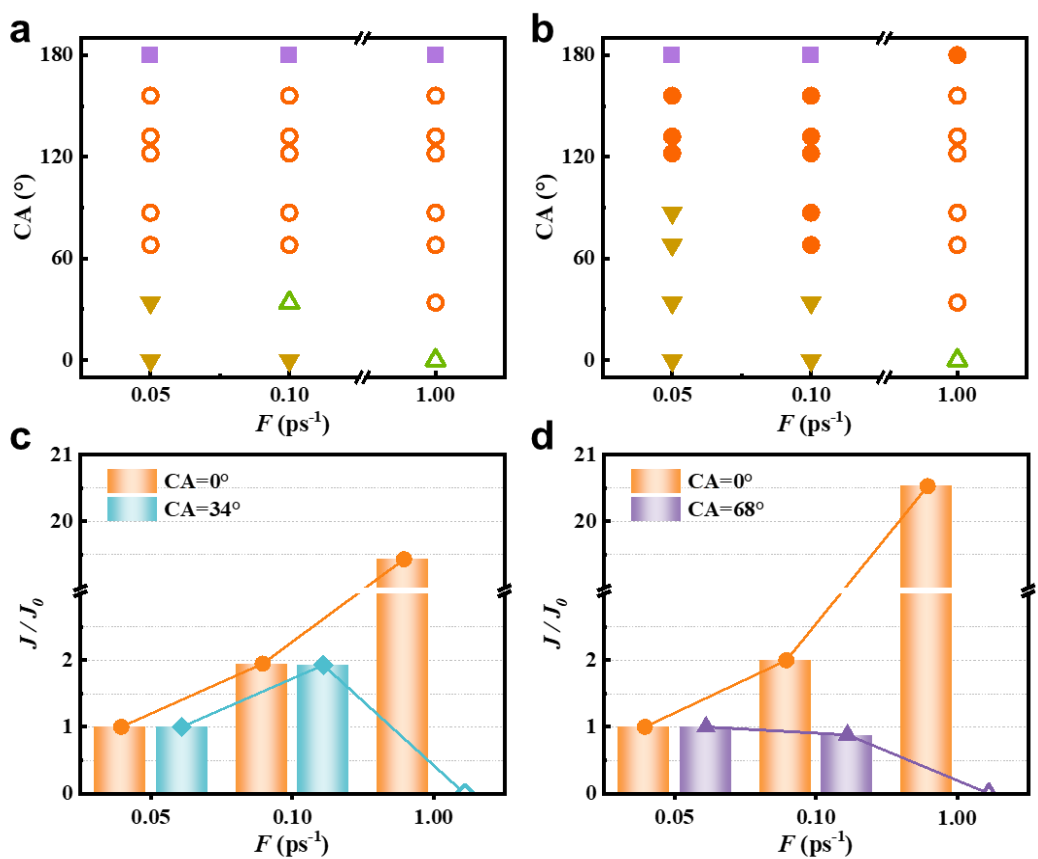

**Figure 5.** Effects of reaction driving force on the nanobubble nucleation and actual reaction rates. (a-b) Phase diagrams of nanobubble nucleation plotted as a function of electrode wettability (i.e., CA) and reaction driving force (characterized by $F$) for (a)

planar- and (b) nano-electrodes. The symbols follow the same conventions as those in Figure 3. (c-d) Relative current density ($J/J_0$) as a function of $F$ at different CAs for (c) planar- and (d) nano-electrodes. $J$ is still set to be 0 when unstable state occurs.

Figure 5 shows the impact of reaction driving force on the bubble nucleation and the relative current density. Here, the relative current density is defined as the ratio of the effective current density at a given $F$ to that at $F = 0.05$ ps$^{-1}$, providing a clearer measure of how increasing the reaction driving force enhances the actual reaction rate. For highly hydrophilic electrodes that are not prone to forming surface bubbles, increasing the reaction driving force improves the effective current density as long as no surface bubbles are formed. For instance, at $CA = 0°$, the current density on both planar- and nano-electrodes increases by approximately 2-fold and 20-fold, respectively, as $F$ increases from 0.05 to 0.1 and 1.0 ps$^{-1}$ (Figure 5 c and d). However, for moderately hydrophilic electrodes (identified as electrodes with optimized wettability based on the findings in Sec. 3.2), excessively increasing the reaction driving force can lead to rapid gas accumulation on the electrode surface, which can result in the formation of surface nanobubbles. Such surface nanobubbles formation can cause prolonged coverage of the electrode, potentially reducing the current density. For example, at $CA = 34°$, the current density on planar electrodes initially increases with $F$ but may subsequently drops significantly due to the formation of unstable surface bubbles (Figure 5 a and c). Similarly, for nanoelectrodes at $CA = 68°$, increasing $F$ from 0.05 to 0.1 ps$^{-1}$ leads to the formation of stable nanobubbles, which prevents an increase in current density but a slight decrease (Figure 5 b and d).

For cases where surface bubbles already form at low reaction driving forces (e.g., planar electrodes with $CA > 60°$), increasing $F$ can accelerate bubble growth (Figure S5) and may promote their detachment, which, from this perceptive, may be favorable for enhancing current densities. For nanoelectrodes where stable surface nanobubbles are initially present, increasing the reaction driving force can destabilize these bubbles, causing them to grow and detach, also potentially enhancing the effective current density. However, the extent of this enhancement is highly dependent on the dynamics of bubble growth, coalescence, and detachment—processes that go beyond the scope of our current simulations and require further detailed investigation.

These findings at different reaction driving forces highlight the need for careful optimization when increasing the applied voltage to enhance the reaction driving force. While faster reactions can significantly improve current density if surface nanobubble formation is avoided, excessively high reaction driving force can cause rapid gas accumulation near the electrode, leading to surface nanobubble formation and adverse effects. Such conditions also introduce more complex bubble dynamics, further influencing the reaction efficiency and requiring a more comprehensive understanding of their impact.

## 3. Conclusions

In this work, we employed molecular dynamics simulations to investigate bubble nucleation and actual reaction rates during water electrolysis on both planar- and nano-electrodes at the nanoscale. By systematically varying the electrode wettability modulated through water-electrode (W-E) and gas-electrode (G-E) interactions, we identified distinct nanobubble nucleation regimes: gas layers, surface nanobubbles, bulk nanobubbles, and no nanobubbles. In general, gas layers form on highly hydrophobic electrodes due to relatively strong G-E attractions, which traps gas molecules on the surface, resulting in complete coverage by a dense gas layer. Surface nanobubbles emerge on moderately hydrophobic electrodes, while bulk nanobubbles are observed on moderately hydrophilic electrodes that no longer promote heterogeneous nucleation but still sustain considerable gas production rates. On highly hydrophilic electrodes, the reduced reactant concentration in the reaction zone diminishes the reaction rate, and insufficient gas supersaturation prevents bubble nucleation, resulting in a no-nanobubble regime.

A key finding is the "volcano-shaped" relationship between the electrode wettability and the actual reaction rate, with the current density peaking at an optimal wettability. This trend is particularly evident under weaker G-E interactions, where an appropriate W-E interaction achieves optimal wettability by suppressing surface nanobubble formation, preventing coverage of reactive areas, and maintaining sufficient reactant concentration in the reaction zone to sustain high reaction probability. Although the underlying mechanisms differ, this relationship parallels the Sabatier Principle in catalytic reaction theory[51-53], where excessively strong or weak interactions between reactants and the catalyst (or electrode) reduce reaction efficiency. Thus, optimal wettability balances nanobubble nucleation and reactant availability, maximizing reaction rates while mitigating adverse effects from gas layers or surface nanobubbles.

Our simulations also reveal a pronounced edge effect on nanoelectrodes, where reaction probabilities are higher near the electrode boundaries. This edge effect results in higher effective current densities on nanoelectrodes compared to planar electrodes under identical wettability conditions. These findings highlight the potential of nanostructured electrodes to significantly improve the utilization efficiency of active materials, particularly in electrochemical reactions involving costly noble metal catalysts[54-56]. This concept is analogous to single-atom catalysis, where maximizing active site utilization is key[57-60]. Our ongoing research indicates that nanoarrayed electrodes exhibit diverse bubble nucleation behaviors, playing a critical role in regulating actual reaction efficiency and offering promising directions for future optimization.

It is important to note that the results presented in this study are based on a simplified molecular simulation model, particularly in terms of the representation of chemical reaction processes. The observed "volcano-shaped" relationship between wettability and current density may vary depending on the underlying reaction mechanism, as it is influenced by how wettability impacts reaction kinetics in the absence of bubbles. Our findings emphasize the need for future simulations to incorporate more realistic reaction mechanisms to better understand how electrode properties affect nanobubble behavior and actual reaction rates. Additionally, due to the size limitations of our simulations, we were unable to fully explore the effects of unstable bubble growth, coalescence, and detachment on actual reaction rates. We simply assumed that prolonged electrode coverage by bubbles in these cases would result in very low current densities, which limits our ability to comprehensively compare reaction performance across varying wettability conditions. These limitations underscore the importance of developing multi-scale models that integrate atomic-scale reaction kinetics, nanoscale bubble nucleation, and microscale bubble dynamics such as growth, coalescence, and detachment. Such an approach would provide a more holistic understanding of bubble behavior and its impact on gas-evolving reactions, offering a pathway toward more efficient and optimized electrode designs—a highly challenging task yet promising direction for future theoretical and computational research.

## 4. Methods

The model used in this work closely follows that in Ref.[38] for the water electrolysis. We employed simulation boxes with dimensions of $10 \times 10 \times H$ nm$^3$, where $H$ represents the box height, which fluctuates under a given pressure, as shown in Figure 1. Water and gas molecules are confined between two solid walls, both constructed from an FCC lattice with a unit cell size of 5 Å and exposing its [001] face to the fluid. The bottom wall served as the electrode, and two types of electrodes were examined: a planar electrode with a surface-wide reaction zone composed of Pt-like atoms (Figure 1 a) and a nanoelectrode with a 6 nm diameter reaction zone surrounded by inert Si-like atoms (Figure 1 b).

The Stillinger-Weber (SW) potential[61] was employed to model interactions between all species, including Pt-like atoms (E), inert Si-like atoms, water/reactant molecules (W), and gas/product molecules (G), with the parameters listed in Table I. The interaction parameters for W-Si were set to represent a highly hydrophilic solid surface for water. The values of $\varepsilon$ for the interactions of W-E and G-E were varied to explore the effect of wettability on the nanobubble nucleation. The contact angle (CA) of an equilibrium nanobubble on the corresponding electrode was measured and is provided in Figure S1. As expected, increasing $\varepsilon_{WE}$ enhances the hydrophilicity of the electrode, while increasing $\varepsilon_{GE}$ makes the electrode more hydrophobic. However, establishing a simple quantitative relationship between CA and the interaction parameters is challenging. Thus, we conducted a series of simulations under various combinations of $\varepsilon_{WE}$ and $\varepsilon_{GE}$ to systematically explore their effects on nanobubble nucleation and reaction rates.

**Table I.** Interaction parameters for the SW potentials between different species. $\varepsilon$ is the depth of the attraction well, and $\sigma$ is the characteristic size. All other parameters are the same as that in Ref.[38].

| type | type | $\varepsilon$ (kcal/mol) | $\sigma$ (Å) |
|---|---|---|---|
| W | W | 6.189 | 2.3925 |
| G | G | 0.14 | 4.08 |
| Pt (i.e., E) | Pt (i.e., E) | 6.189 | 3.56 |
| Si | Si | 6.189 | 3.56 |
| W | G | 0.18 | 4.00 |
| W | Si | 1.00 | 3.56 |
| W | Pt (i.e., E) | 0.20~1.00 | 3.56 |
| G | Si | 0.20 | 4.08 |
| G | Pt (i.e., E) | 0.20~1.00 | 4.08 |

All simulations were carried out using the open-source code LAMMPS[62-63] in the isothermal-isostress ($NP^{zz}T$) ensemble, with 34400 fluid molecules (including both water and gas), 3200 atoms in both bottom and top walls. The mass of water, gas, Si-like, and Pt-like particles are 18, 2, 28 and 195 amu, respectively. The inert top wall was subjected to an external force along the z-direction and moved as a rigid body to maintain a pressure of $p = 1$ atm. During all simulations, atoms in the bottom wall were constrained at their initial positions through soft harmonic potentials with $k = 10$ kcal·mol$^{-1}$·Å$^{-2}$ to preserve its structural integrity. Periodic boundary conditions were used in the x and y directions. The velocity Verlet algorithm, with a time step of 10 fs, was used to integrate of equations of motion. The Nosé-Hoover thermostat with a time constant of 0.5 ps was used to control the temperature of the fluid and bottom wall at $T = 300$ K. Each simulation box has a gas reservoir (Figure 1), where all gas molecules are converted into water molecules every 0.01 ns to ensured that the gas concentration far from nanobubbles remained at 0.

To mimic the electrochemical reaction process, water molecules near Pt-like atoms were converted into gas molecules at fixed simulation intervals. This conversion only occurs for water molecules within a distance ≤ 2.5 Å from a Pt-like atom, with each Pt-like atom capable of catalyzing only one water molecule at a time. The frequency $F$ of conversion attempts represent *the reaction driving force*, and high values of $F$ corresponded to conditions analogous to applying a high electric voltage to the electrode that can lead to a high intrinsic reaction rate. The simulation system was pre-equilibrated for at least 1 ns before activating the reaction algorithm. Production runs were collected for durations of up to 100 ns or until no new gas molecules were produced within 10 consecutive reaction periods.

**Declaration of competing interest**

The authors declare that they have no known competing financial interests or personal relationships that could have appeared to influence the work reported in this paper.

**Acknowlegments**

The work was supported by National Natural Science Foundation of China (22208343, U22A20416), Beijing Natural Science Foundation (2242020), CAS Project for Young Scientists in Basic Research (YSBR-105) and State Key Laboratory of Mesoscience and Engineering (MESO-23-A09).

# SI for
# Tuning Electrode Wettability to Optimize Nanobubble Nucleation and Reaction Rates in Electrochemical Gas-Evolving Reactions

Zhenlei Wang[a,b,c], Yaxi Yu[a,b,c], Mengkai Qin[a,b,c], Hao Jiang[a,b,c], Zhenjiang Guo[b], Lu Bai[a], Limin Wang[b,c], , Xiaochun Zhang[a,b] *, Xiangping Zhang[a,b] *, Yawei Liu[a,b,c,d] *

[a]*Beijing Key Laboratory of Solid State Battery and Energy Storage Process, CAS Key Laboratory of Green Process and Engineering, Institute of Process Engineering, Chinese Academy of Sciences, Beijing 100190, P. R. China*

[b]*State Key Laboratory of Mesoscience and Engineering, Institute of Process Engineering, Chinese Academy of Sciences, Beijing 100190, P. R. China*

[c]*University of Chinese Academy of Sciences, Beijing 100049, P. R. China*

[d]*Longzihu New Energy Laboratory, Zhengzhou Institute of Emerging Industrial Technology, Henan University, Zhengzhou 450000, P. R. China*

*E-mail: xchzhang@ipe.ac.cn, xpzhang@ipe.ac.cn, ywliu@ipe.ac.cn

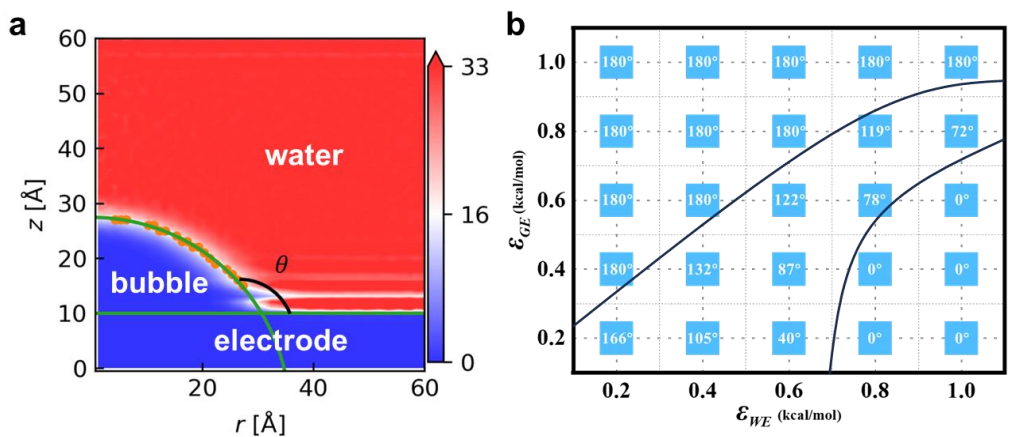

**Figure S1.** (a) Schematic illustration of the measurement of bubble contact angles on the electrode surface, where the contact angle is defined as the angle formed by the outer liquid phase. The interface, represented by the green curve, is defined as the region where the liquid density is half of the bulk density. (b) Contact angles under various W-E and G-E interactions, reflecting the wettability characteristics of different electrodes.

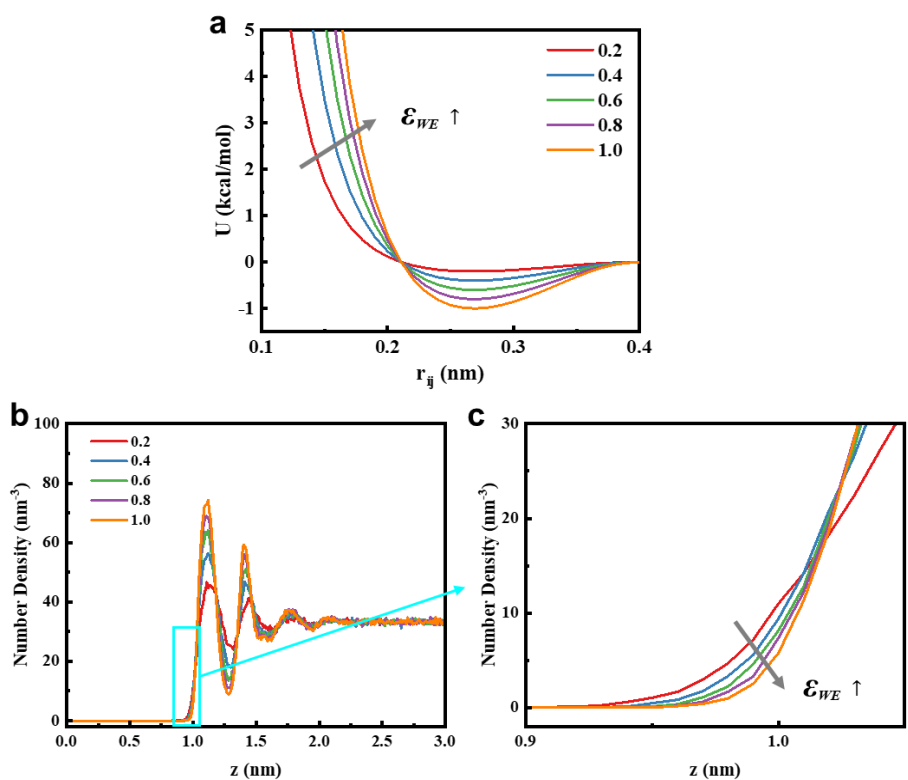

**Figure S2**. Profiles of water/reactant number density at the electrode interface under varying water-electrode interactions ($\varepsilon_{WE}$). It is evident that as $\varepsilon_{WE}$ decreases,

indicating a more hydrophobic electrode, the water density within the reaction zone (i.e., the region within 0.25 nm from the electrode surface) becomes significantly lower.

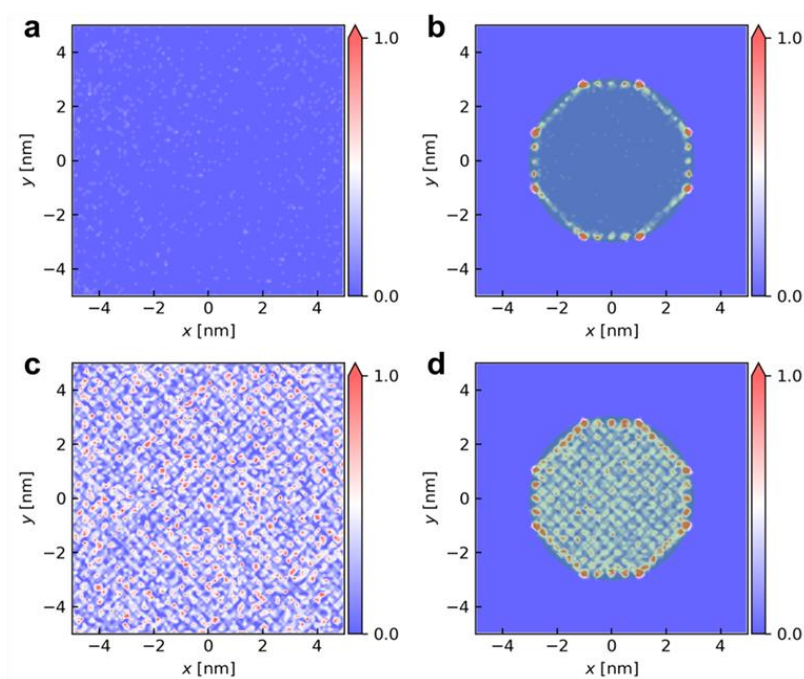

**Figure S3**. Density distributions of reaction event occurring on (a) the planar electrode plane with a gas layer, (b) the nanoelectrode with a surface nanobubble, (c) the planar electrode with no bubbles and (d) the nanoelectrode with no bubbles.

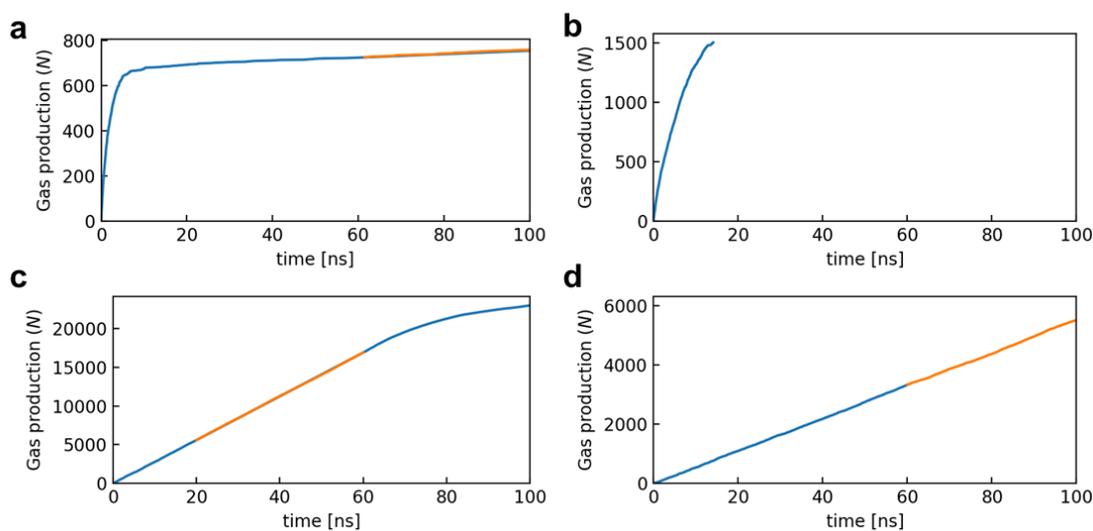

**Figure S4.** The number of gas molecule production as a function of time on planar electrodes with (a) a gas layer, (b) an unstable surface nanobubble, (c) a bulk nanobubble, and (d) no nanobubbles. The orange line indicates the slope of the curve

in the steady state, which is measured to represent the actual gas production per unit time. For the unstable surface nanobubble, there is no steady state.

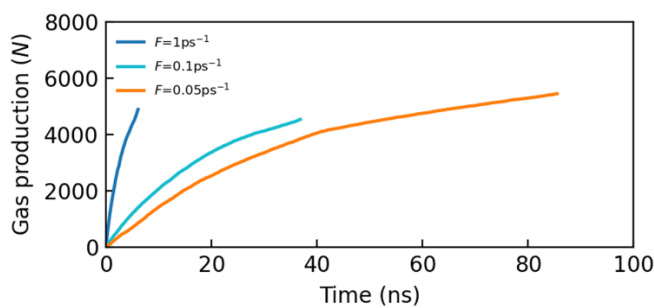

**Figure S5.** The number of gas molecules produced over time under different reaction driving forces. The termination of each curve indicates that uncontrolled bubble expansion eventually halts the reaction. At higher reaction driving forces, the curves terminate earlier, reflecting faster bubble growth.

**Movie 1.** Videos for nanobubble nucleation on (a-d) planar electrodes and (e-h) nanoelectrodes at varying wettabilities with (a, e) CA~180° ($\varepsilon_{WE}, \varepsilon_{GE} = 0.6, 1.0$ kcal/mol), (b, f) CA~122° ($\varepsilon_{WE}, \varepsilon_{GE} = 0.6, 0.6$ kcal/mol), (c, g) CA~40° ($\varepsilon_{WE}, \varepsilon_{GE} = 0.6, 0.2$ kcal/mol), and (d, h) CA~0° ($\varepsilon_{WE}, \varepsilon_{GE} = 1.0, 0.2$ kcal/mol).

**Movie 2.** Videos for the fluid distributions in the reaction zone of (a) the planar electrode covered by a gas layer and (b) the nanoelectrode covered by a surface nanobubble. It reveals that thermodynamic fluctuations at the boundaries of the nanoelectrode intermittently expose the electrode surface, enhancing the reaction probability on the electrode.